\documentstyle[12pt,twoside,fleqn,espcrc1]{article}
\input epsf
\newcommand{\AmS}{{\protect\the\textfont2
  A\kern-.1667em\lower.5ex\hbox{M}\kern-.125emS}}

\title{
Scaling of particle production with number of participants
in high-energy A+A collisions in the parton-cascade model}

\author{Dinesh Kumar Srivastava\address{Variable Energy Cyclotron
Centre,\\ 1/AF Bidhan Nagar, Calcutta 700 064, India}
        and 
Klaus Geiger\address{Physics Department, Brookhaven National
Laboratory, Upton,\\
New York 11973, U. S. A.}}

\begin{document}
% typeset front matter
\maketitle

\begin{abstract}

In view of the recent WA98 data of $\pi^0$ spectra from
central $Pb+Pb$ collisions at the CERN SPS, we analyze
the production of neutral pions for
$A+A$ collisions  across the periodic table at
 $\sqrt{s}=17$ AGeV and 200 AGeV within the framework of
 the parton-cascade  model for relativistic
heavy ion collisions. The multiplicity of the pions (having $p_T > 0.5$ GeV)
in the central rapidity region, is
seen to scale as $\sim (N_{part})^{\alpha}$,
where $N_{part}$ is the number of participating nucleons, which we
have approximated as $2A$ for central collisions of identical
nuclei.  We argue that the deviation of $\alpha$ ($\simeq 1.2$) from unity
 may have its origin in the multiple scattering suffered by the partons.
We also find that the constant of proportionality in the above
scaling relation increases substantially in going from
SPS to RHIC energies.  This would imply that the (semi)hard partonic
activity becomes a much cleaner signal above
the soft particle production at the higher energy of RHIC, and thus
much less dependent on the (lack of)
understanding of the underlying soft physics background.
\end{abstract}

\section*{}
Recently, the WA98 Collaboration has published \cite{wa98}
 data for the production of neutral pions up to
transverse momenta of $p_\perp \simeq 4$ GeV/c,
in central $Pb+Pb$ collisions at 160 A GeV/c incident momentum,
corresponding to $\sqrt{s}\simeq 17$ A GeV.
Two most interesting features of these data emerge when
compared to corresponding data from $pp$ collisions and
collisions involving lighter nuclei \cite{wa98}:
a) an approximate invariance of the spectral shapes, i.e.,
a near indepence of the slope of the neutral pion $p_\perp$ spectra;
b) a simple scaling of the $\pi^0$ with the number
of participating nucleons, if the number of participants is large
($ \,\lower3pt\hbox{$\buildrel > \over\sim$}\,30$).
In the present work,
we use the event generator VNI \cite{vni} which embodies the physics
of the parton-cascade model~\cite{pcm} for 
ultra-relativistic heavy-ion collisions to analyze these observations.
The model attempts to describe the nuclear dynamics
on the microscopic level of particle transport and interactions,
by evolving the multi-particle system in space-time from the instant
of nuclear overlap all the way to the final-state hadron yield. 
For details we refer the interested reader to Refs. \cite{vni,pcm}.

A simple consideration may illustrate the features
of particle production within this approach and its relevance to
multiple parton scattering. Let $x$ denote the number of
partons in each nucleon, and let each parton suffer $\nu$ collisions
during the partonic stage. Assuming that each virtual parton radiates
$r$ partons, we see that the number of produced partons will
vary as
\begin{equation}
N_{\mbox{\footnotesize{partons}}}\,\;\propto\,\; \nu \; (1+r) \;\, x \;\, A 
\;,
\end{equation}
and as we expect, 
$
N_{\mbox{\footnotesize{hadrons}}}\,\;\propto\,\;
   N_{\mbox{\footnotesize{partons}}}
\;,
$
one realizes immediately that if the partons interact only once, 
the multiplicity of the partons, and hence the multiplicity of hadrons, 
will scale as $A$. It is also clear that if every parton 
interacts with with every other
parton then $\nu \propto A$, and the number of materialized partons 
would scale as $A^2$. That can happen, if the system would live for an 
infinitely long time.
However, this is not the case. Rather than that, in 
relativistic heavy ion collisions, the partonic matter will expand,
dilute, and eventually convert into hadrons. Thus a given parton may
undergo $\nu \sim R/ \lambda$ 
interactions; where $R$ is the transverse 
size of the system and $\lambda$ is the mean free path of the parton.
Noting that $R\sim A^{1/3}$,  we immediately see that the
number of materialized partons, and hence the number of produced particles
would scale as $\approx \, A^{4/3}$. An experimental verification of this
scaling behaviour could be a direct manifestation the formation
of a  dense partonic matter!

We shall demonstrate now that these simple considerations are 
indeed confirmed by a detailed simulation
with the event generator VNI on the basis of the 
parton-cascade/cluster-hadronization model.
We first consider the recently measured transverse momentum
distribution of $\pi^0$-production in central collsions of $Pb+Pb$
at CERN SPS obtained by the WA98 collaboration~\cite{wa98}. 
To make contact with the experimental data, 
the simulations were done
for the range of impact parameters $0 < b < 4.5 $ fm, which corresponds to 10\% 
of minimum-bias cross-section. The result of our  model calculation,
shown as the solid histogram in Fig. 1a, is seen to be in decent agreement with 
the experimental measurements.
The model results do not include the final-state interaction
among produced hadrons yet, but it is likely \cite{ron} that the agreement will 
further improve once the effect of cascading hadrons is included. 
The dashed histogram in Fig. 1 gives the soft contribution while the
solid curve gives a hydrodynamic prediction (without the contribution
of resonance decays).

In Fig. 2 we plot our results for the $p_\perp$ spectra of $\pi^0$'s for 
a number of central $AA$ collisions at $\sqrt{s}=$ 17 A GeV for
various $A+A$ systems from $A=16$ to  $A= 197$. One observes that
they are almost identical in shape with a universal
 slope for 
$p_\perp \,\lower3pt\hbox{$\buildrel < \over\sim$}\,1.5$ GeV/c.
On  the other hand, the deviations appearing at larger $p_\perp$ for heavier
systems are indicative of enhanced multiple scattering there.
Similar results (not displayed here) were obtained at RHIC energies.
In order to verify 
this scaling more closely, we have calculated, as a function of the nuclear
mass number $A$, the production of $\pi^0$'s
in the central rapidity region ($-0.5 < y < 0.5$)
having transverse momenta $p_\perp \ge 0.5$ GeV/c. The latter choice minimizes
the influence of pions having their origin in decay of resonances. This
kinematic window was motivated \cite{wa98} by the WA98 collaboration
in their measurement of the $\pi^0$ yield.
Fig. 3 displays the simulation results for central $A+A$
collisions across the periodic table, at CERN SPS center-of-mass energy
$\sqrt{s} = 17 $ A GeV, while Fig. 4 shows the same for RHIC energy 
$\sqrt{s} = 200 $ A GeV.  The solid lines are  fits to the model results,
represented by the symbols, and scale as 
%%%%%%%%%%%%%%%%%%%%%%%%%%%%%%%%  Fig. 1,2 %%%%%%%%%%%%%%%%%%%%%%%%%%
\setcounter{figure}{0}
\begin{figure}[htb]
\begin{minipage}[t]{80mm}
\epsfxsize=3.25in
\epsfbox{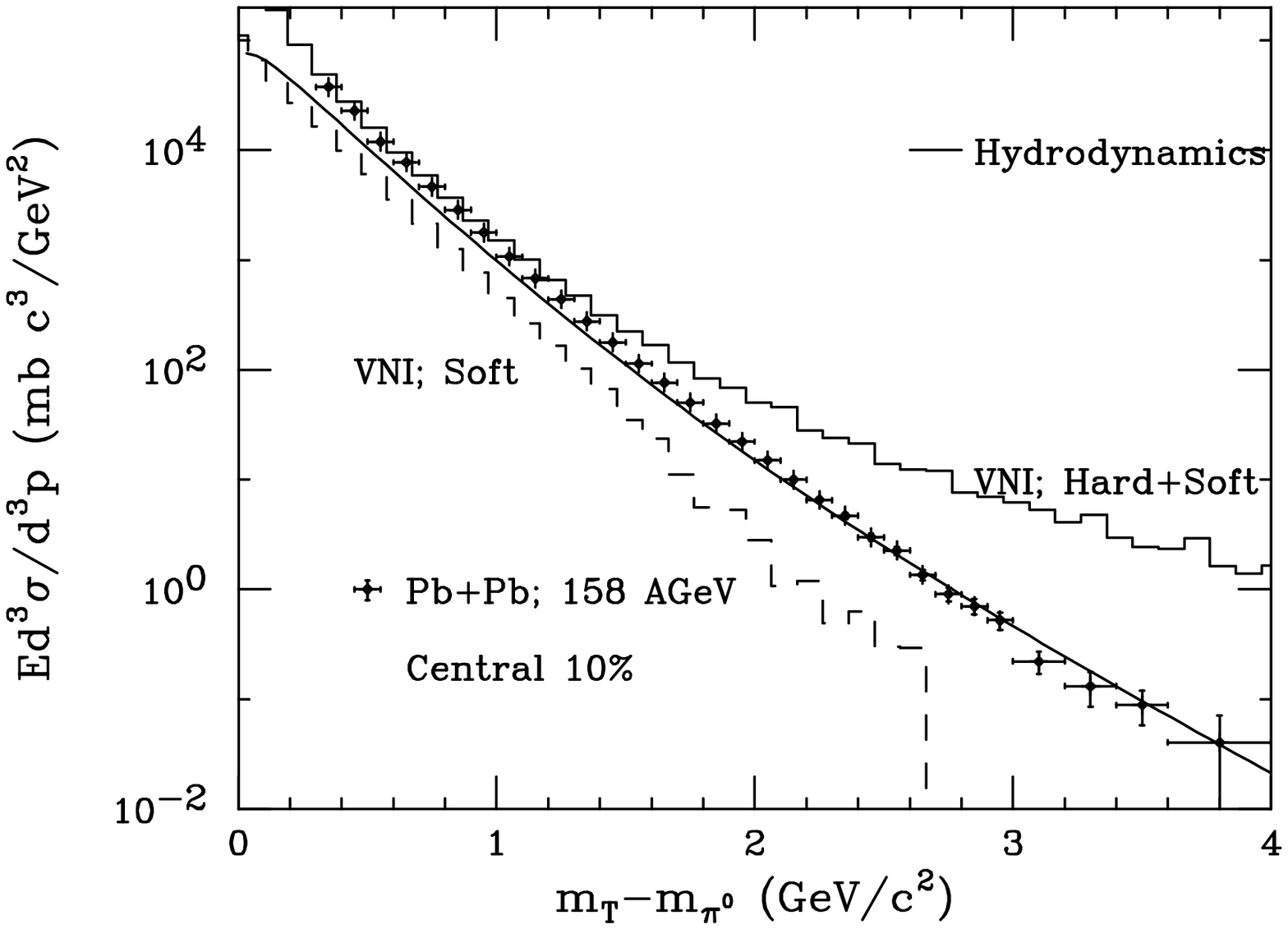}
\label{wa98}
\end{minipage}
\hspace{0.04\linewidth}
\begin{minipage}[b]{0.37\linewidth}
\caption{ Transverse mass spectra of neutral pions in central
collision of 158 AGeV $Pb+Pb$ collisions. The solid histogram 
represents our result from the parton-cascade/cluster-hadronization 
model. The dashed histogram gives the contribution of the soft-part.}
\end{minipage}
\begin{minipage}[t]{80mm}
\epsfxsize=3.25in
\epsfbox{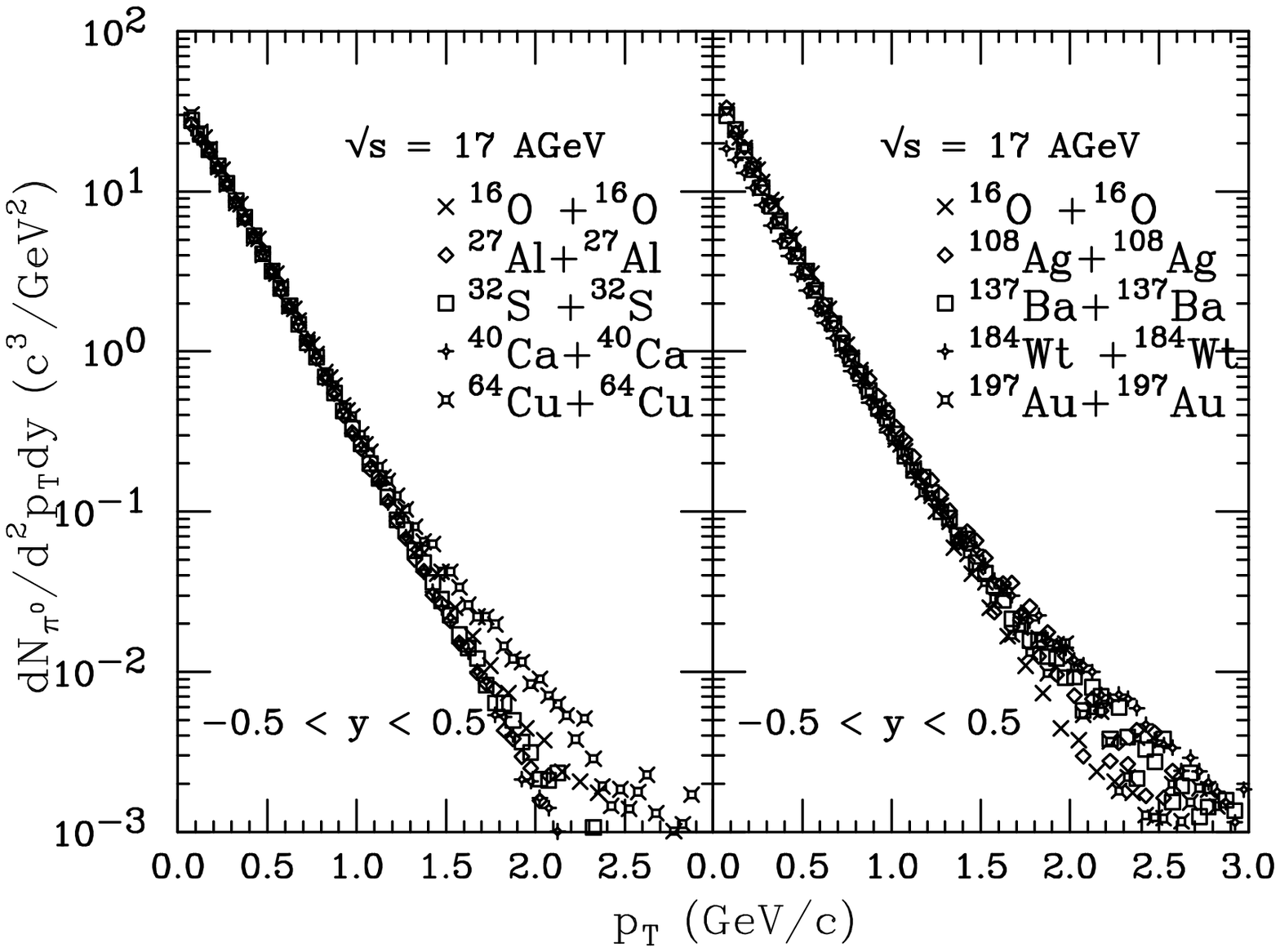}
\label{all}
\end{minipage}
\hspace{0.04\linewidth}
\begin{minipage}[b]{0.35\linewidth}
\caption{
Transverse momentum spectra of neutral pions in central
collisions of identical nuclei at $E_{cm} = 17$ A GeV.
The symbols are our results for various systems, from $O+O$ up to $Au+Au$.
All results  are normalized to the case  $A=16$.
}
\end{minipage}
\end{figure}
\vspace{-2mm}
%%%%%%%%%%%%%%%%%%%%%%%%%%%%%%%%  Fig. 1,2 %%%%%%%%%%%%%%%%%%%%%%%%%%
\begin{equation}
N_{\pi^0} \;\propto\; \left( N_{part}\right)^{\alpha} 
\label{prop}
\;,
\end{equation}
where $N_{part}=2A$ is the number of participating nucleons, and
$\alpha$ being extracted as:
\begin{equation}
\alpha
\; \approx \;
\left\{
\begin{array}{l}
1.16 \;\;\; \mbox{at}\;\;\sqrt{s}=17 \;\mbox{A GeV} \\
1.23 \;\;\; \mbox{at}\;\;\sqrt{s}=200 \;\mbox{A GeV} 
\label{alpha}
\end{array}
\right.
\label{prop1}
\;.
\end{equation}
It is interesting to note that
$\alpha \approx 1.2$ is in excellent agreement with the
corrected WA98 results \cite{remark2}.

%%%%%%%%%%%%%%%%%%%%%%%%%%%%%%%%  Fig. 3 4 %%%%%%%%%%%%%%%%%%%%%%%%%%
\setcounter{figure}{2}
\begin{figure}[htb]
\begin{minipage}[t]{77mm}
\epsfxsize=3.in
\epsfbox{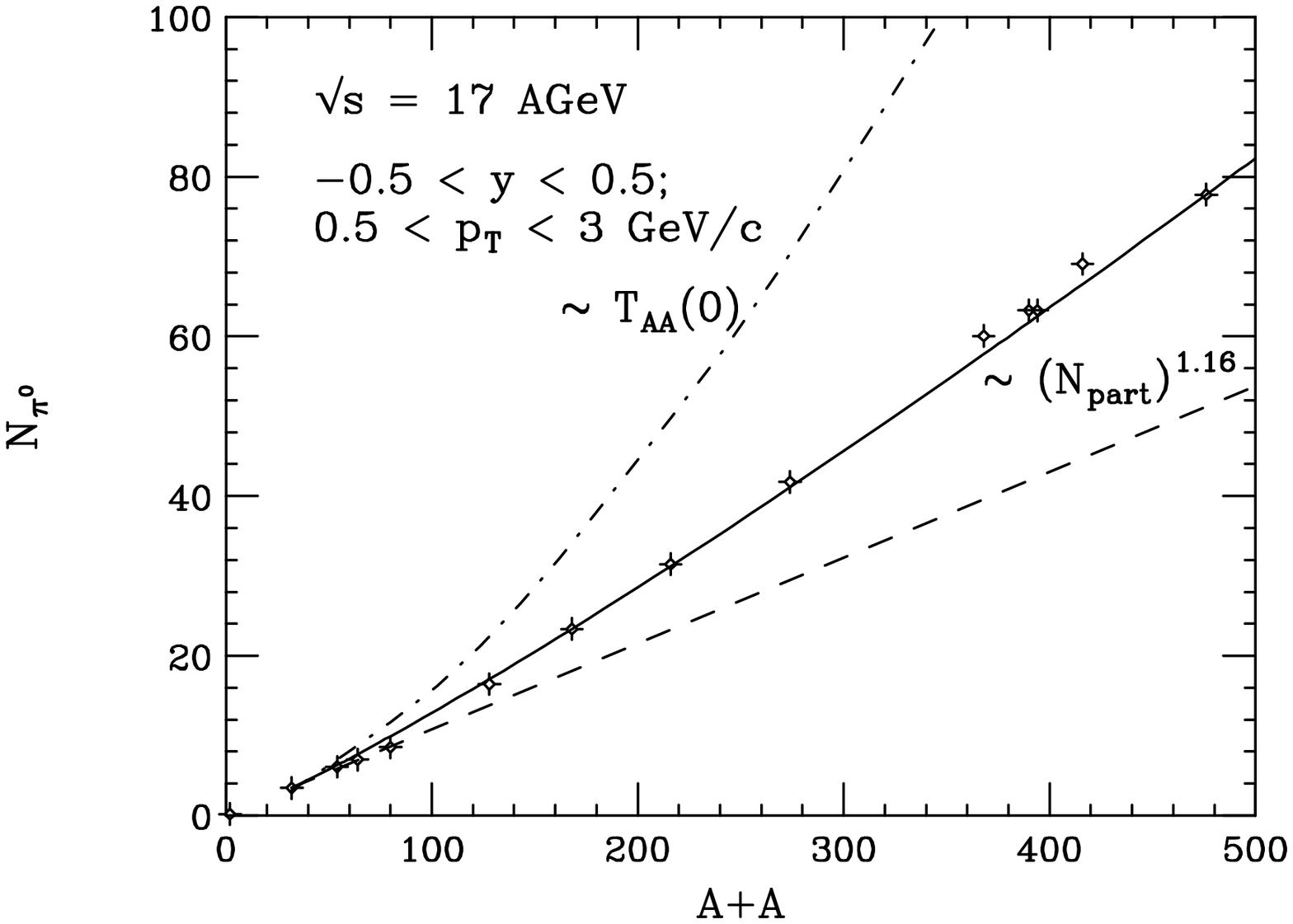} 
\label{sps}
\end{minipage}
\hspace{0.04\linewidth}
\begin{minipage}[b]{0.37\linewidth}
\caption{
Mass number scaling of pion production in the central rapidity region
at CERN SPS energies.
 The symbols represent the results of
our simulations, and the solid curve is a fit to these results.}
\end{minipage}

\begin{minipage}[t]{77mm}
\epsfxsize=3.in
\epsfbox{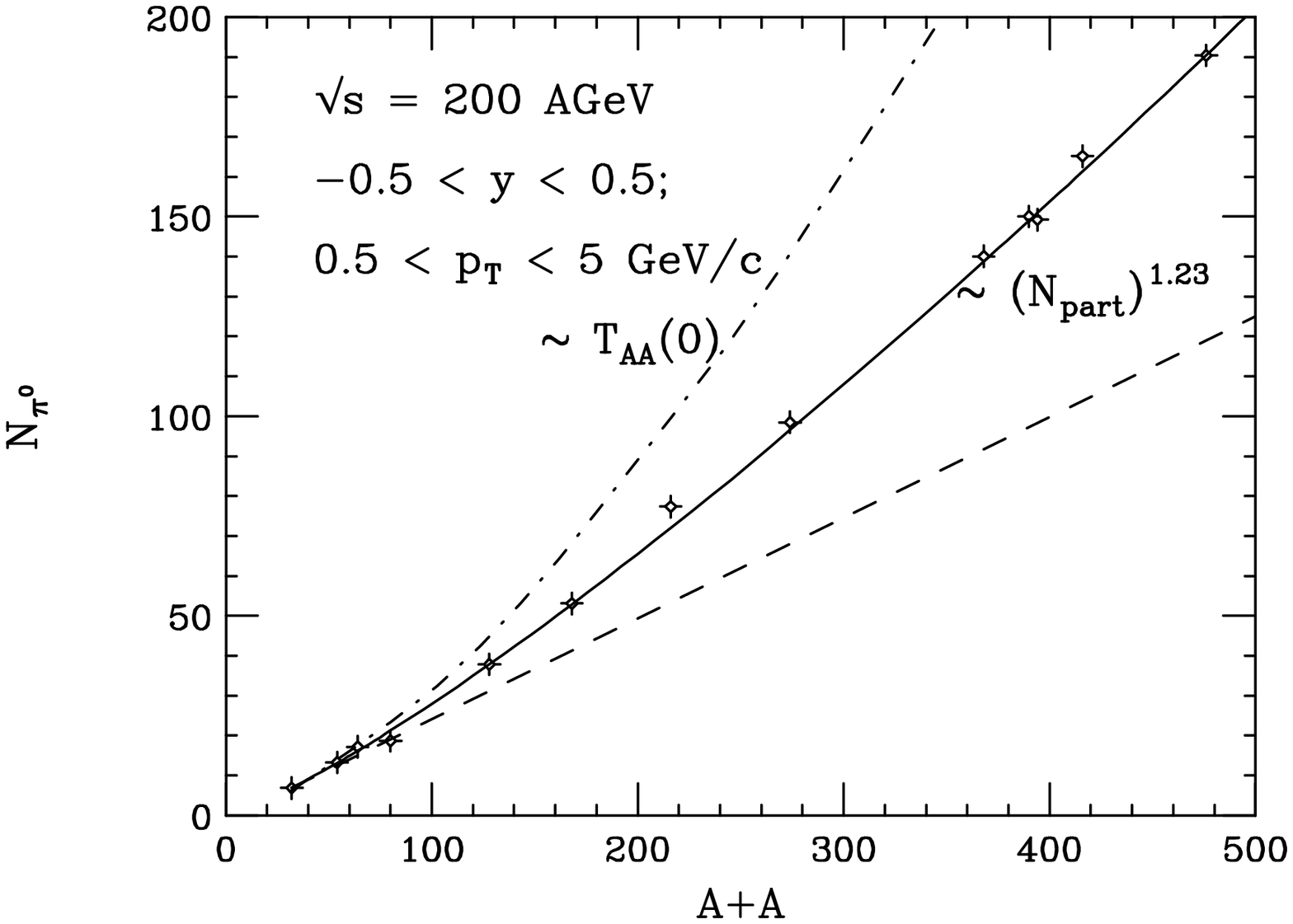}
\label{rhic}
\end{minipage}
\hspace{0.04\linewidth}
\begin{minipage}[b]{0.37\linewidth}
\caption{ Same as above for BNL RHIC energies. Both the increased constant
of proportionality as well the exponent for scaling imply an increased
multiple scattering at the (higher) RHIC energies.}
\end{minipage}
\end{figure}
%%%%%%%%%%%%%%%%%%%%%%%%%%%%%%%%  Fig. 2 a,b %%%%%%%%%%%%%%%%%%%%%%%%%%

For comparison, the dashed lines correspond to a linear scaling
$ \sim ( N_{part})^{1.0}$, whereas the dashed-dotted
lines indicate a hypothetical scaling with the nuclear overlap factor
$T_{AA}(b=0)\sim (N_{part})^{1.42}$. 
A linear scaling would reflect a single-collision situation,
and a scaling with $T_{AA}$ would indicate a Glauber-type multiple-collision 
scenario, in which nucleons suffer several collisions
along their incident straight-line trajectory, without deflection
and without energy loss.
Comparing the three curves, we can conclude that
our simulation results rise significantly slower than
with $T_{AA}$, because firstly, the particles change direction through the
collisions, and secondly, they are subject to a collision time of the order
of the inverse momentum transfer, during which they cannot rescatter.
On the other hand, our calculated $\pi^0$ yields grow much faster than
linear with $A$, due to multiple scatterings.
\bigskip

In summary, we have demonstrated here that the observed scaling  
of the number of produced particles with the number of participants,
 in heavy-ion $A+A$ collisions, as well as
the approximate shape-independence of the transverse momentum spectra, 
are satisfactorily reproduced by the parton-cascade / cluster-hadronization
model. We must add that a more accurate description of the $p_T$ spectra
at larger $p_T$ will need a readjustment of the hadronization model.

{\em {Dr. Klaus Geiger, a dear colleague, a very good friend, and a brilliant
physicist was killed in an air-crash in September 1998. This is one of
the last pieces of work which we completed together~\cite{web}.}} 

%\newpage


\begin{thebibliography}{9}
\bibitem{remark2}
 The WA98 group reported at the QM'99 meeting (see talk by T. Peitzmann)
that the value of $\alpha$ reported by them in Ref.~\cite{wa98} $\approx$ 1.3
was now corrected to $\approx$ 1.15--1.20. This is in much better
agreement with our prediction which were made before the revised
data were reported.
\end{thebibliography}
\end{document}